\documentclass[11pt]{article}
\usepackage[pdftex]{graphicx}
\usepackage{amsmath}
\usepackage{amssymb}
\usepackage{amsthm}
\usepackage{amsfonts}
\usepackage{cite}
\usepackage{float}
\usepackage{hyperref}
\newcommand{\be}{\begin{eqnarray}}
\newcommand{\ee}{\end{eqnarray}}
\newcommand{\bez}{\begin{eqnarray*}}
\newcommand{\eez}{\end{eqnarray*}}

\newcommand{\cA}{\mathcal{A}}
\newcommand{\cB}{\mathcal{B}}

\newcommand{\bA}{\boldsymbol{A}}
\newcommand{\bB}{\boldsymbol{B}}

\newcommand{\bI}{\boldsymbol{I}}

\newcommand{\bK}{\boldsymbol{K}}
\newcommand{\bM}{\boldsymbol{M}}

\newcommand{\bP}{\boldsymbol{P}}

\newcommand{\bQ}{\boldsymbol{Q}}
\newcommand{\bR}{\boldsymbol{R}}
\newcommand{\bS}{\boldsymbol{S}}

\newcommand{\bT}{\boldsymbol{T}}

\newcommand{\bu}{\boldsymbol{u}}
\newcommand{\bU}{\boldsymbol{U}}
\newcommand{\bv}{\boldsymbol{v}}
\newcommand{\bV}{\boldsymbol{V}}

\newcommand{\bX}{\boldsymbol{X}}
\newcommand{\bXi}{\boldsymbol{\Xi}}
\newcommand{\bY}{\boldsymbol{Y}}

\newcommand{\bbC}{\mathbb{C}}

\newcommand{\tbU}{\tilde{\boldsymbol{U}}}
\newcommand{\tbV}{\tilde{\boldsymbol{V}}}

\renewcommand{\d}{\mathrm{d}}
\newcommand{\bd}{\bar{\mathrm{d}}}
\newcommand{\D}{\mathrm{D}}

\theoremstyle{plain}
\newtheorem{theorem}{Theorem}
\newtheorem{proposition}[theorem]{Proposition}

\date{ }

\begin{document}

\title{\bf Bidifferential calculus, matrix SIT \\ and sine-Gordon equations}

\author{ \sc{Aristophanes Dimakis} \\ 
 Department of Financial and Management Engineering \\
 University of the Aegean, 41 Kountourioti Str., GR-82100 Chios, Greece \\
 E-mail: dimakis@aegean.gr 
   \and
  \sc{Nils Kanning} and \sc{Folkert M\"uller-Hoissen} \\
 Max-Planck-Institute for Dynamics and Self-Organization \\
 Bunsenstrasse 10, D-37073 G\"ottingen, Germany \\
 E-mail: nils.kanning@ds.mpg.de, folkert.mueller-hoissen@ds.mpg.de
}

\maketitle

\begin{abstract}
We express a matrix version of the self-induced transparency (SIT) equations in the 
bidifferential calculus framework. An infinite family of exact solutions is then obtained 
by application of a general result that generates exact solutions from solutions of a 
linear system of arbitrary matrix size. A side result is a solution formula 
for the sine-Gordon equation. 
\end{abstract}

\paragraph{1. Introduction.}
The bidifferential calculus approach (see \cite{DMH08bidiff} and the references therein) aims to 
extract the essence of integrability aspects of integrable partial differential or difference equations 
(PDDEs) and to express them, and relations between them, in a universal way, i.e. resolved from specific examples. 
A powerful, though simple to prove, result \cite{DMH08bidiff,DMH10NLS,DMH10AKNS} (see section~6) 
generates families of exact solutions from a matrix linear system. In the following we briefly recall the 
basic framework and then apply the latter result to a matrix generalization of the SIT equations.

\paragraph{2. Bidifferential calculus.}
A \emph{graded algebra} is an associative algebra $\Omega$ over $\bbC$ with a direct 
sum decomposition $\Omega = \bigoplus_{r \geq 0} \Omega^r$ 
into a subalgebra $\cA := \Omega^0$ and $\cA$-bimodules $\Omega^r$, such that
$\Omega^r \, \Omega^s \subseteq \Omega^{r+s}$.
A \emph{bidifferential calculus} (or \emph{bidifferential graded algebra}) is a unital graded algebra
$\Omega$ equipped with two ($\bbC$-linear) graded derivations
$\d, \bar{\d} : \Omega \rightarrow \Omega$ of degree one (hence $\d \Omega^r \subseteq \Omega^{r+1}$,
$\bar{\d} \Omega^r \subseteq \Omega^{r+1}$), with the properties
\be
    \d_z^2 = 0    \qquad\forall z \in \bbC \, , \qquad 
    \mbox{where} \quad \d_z := \bd - z \, \d \, , 
            \label{bidiff_conds}
\ee
and the graded Leibniz rule $\d_z(\chi \, \chi') = (\d_z \chi) \, \chi' + (-1)^r \, \chi \, \d_z \chi'$,
for all $\chi \in \Omega^r$ and $\chi' \in \Omega$.

\paragraph{3. Dressing a bidifferential calculus.}
Let $(\Omega, \d, \bd)$ be a bidifferential calculus. Replacing $\d_z$ in (\ref{bidiff_conds}) by
$\D_z := \bd - A - z \, \d$ with a 1-form $A \in \Omega^1$ (in the expression for $\D_z$ to be 
regarded as a multiplication operator), the resulting condition $\D_z^2 =0$ (for all $z \in \bbC$) 
can be expressed as 
\be
    \d A = 0  =  \bd A - A \, A  \; .       \label{gauged-bdga-cond}
\ee
If (\ref{gauged-bdga-cond}) is equivalent to a PDDE, we have a 
\emph{bidifferential calculus formulation} for it. This requires that $A$ depends on  
independent variables and the derivations $\d, \bd$ involve differential or difference operators. 
Several ways exist to reduce the two equations (\ref{gauged-bdga-cond}) to a single one: \\
\noindent
(1) We can solve the first of (\ref{gauged-bdga-cond}) by setting $A = \d \phi$.
This converts the second of (\ref{gauged-bdga-cond}) into
\be
     \bd \, \d \, \phi = \d\phi \; \d\phi \; .    \label{phi_eq}
\ee
\noindent
(2) The second of (\ref{gauged-bdga-cond}) can be solved by setting
$A = (\bd g) \, g^{-1}$. The first equation then reads
\be
        \d \, \Big( (\bd g) \, g^{-1}\Big) =0 \; .   \label{g_eq}
\ee
(3) More generally, setting $A = [\bd g - (\d g) \Delta] \; g^{-1}$, 
with some $\Delta \in \cA$, we have 
$\bd A - A \, A = (\d A) \, g \Delta g^{-1} + (\d g) \, (\bd \Delta - (\d \Delta) \, \Delta) \, g^{-1}$.
As a consequence, if $\Delta$ is chosen such that $\bd \Delta = (\d \Delta) \, \Delta$, 
then the two equations (\ref{gauged-bdga-cond}) reduce to
\be
      \d \Big( [\bd g - (\d g) \Delta] \; g^{-1} \Big) = 0 \; .  \label{g_mod_eq}
\ee

With the choice of a suitable bidifferential calculus, (\ref{phi_eq}) and (\ref{g_eq}), or more 
generally (\ref{g_mod_eq}), have been shown to reproduce quite a number of integrable PDDEs. 
This  includes the self-dual Yang-Mills equation, in which case (\ref{phi_eq}) 
and (\ref{g_eq}) correspond to well-known potential forms \cite{DMH08bidiff}. 
Having found a bidifferential calculus 
in terms of which e.g. (\ref{phi_eq}) is equivalent to a certain PDDE, it is not in general 
guaranteed that also (\ref{g_eq}) represents a decent PDDE. Then the generalization (\ref{g_mod_eq}) 
has a chance to work (cf. \cite{DMH08bidiff}). In such a case, the \emph{Miura transformation}
\be
     [\bd g - (\d g) \Delta] \; g^{-1} = \d \phi   \label{Miura}
\ee
is a hetero-B\"acklund transformation relating solutions of the two PDDEs. 

B\"acklund, Darboux and binary Darboux transformations can be understood  
in this general framework \cite{DMH08bidiff}, and there is a construction of an infinite set of 
(generalized) conservation laws. Exchanging $\d$ and $\bd$ leads  
to what is known in the literature as `negative flows' \cite{DMH10AKNS}. 

\paragraph{4. A matrix generalization of SIT equations and its Miura-dual.}
$\cA = \mathrm{Mat}(n,n,C^\infty(\mathbb{R}^2))$ denotes the algebra of $n \times n$ matrices of smooth 
functions on $\mathbb{R}^2$. Let $\Omega = \cA \otimes \bigwedge(\mathbb{C}^2)$ with 
the exterior algebra $\bigwedge(\mathbb{C}^2)$ of $\mathbb{C}^2$.
In terms of coordinates $x,y$ of $\mathbb{R}^2$, a basis $\zeta_1, \zeta_2$ of $\bigwedge^1(\mathbb{C}^2)$, 
and a constant $n \times n$ matrix $J$, maps $\d$ and $\bd$ are defined as follows on $\cA$,
\begin{align*}
     \d f = \tfrac{1}{2} [J,f] \otimes \zeta_1 + f_y \otimes \zeta_2 \, , \qquad
    \bd f = f_x \otimes \zeta_1 + \tfrac{1}{2} [J,f] \otimes \zeta_2 
\end{align*}
(see also \cite{Gris+Pena03}). 
They extend in an obvious way (with $\d \zeta_i = \bd \zeta_i =0$) to $\Omega$ such that $(\Omega,\d,\bd)$ 
becomes a bidifferential calculus. We find that (\ref{phi_eq}) is equivalent to
\begin{align}  \label{phi_eq_1}
  \phi_{xy}=\tfrac{1}{2}\left[[J,\phi],\phi_y - \tfrac{1}{2}J\right] \; .
\end{align}
Let $n=2 m$ and $J = \mbox{\rm block-diag}(I, -I)$, where $I=I_m$ denotes the $m \times m$ identity matrix. 
Decomposing $\phi$ into $m \times m$ blocks, and constraining it as follows,
\begin{align} \label{phi_matrixSIT}
  \phi=
  \begin{pmatrix}
    p&q\\
    q&-p\\
  \end{pmatrix} \, ,
\end{align}
(\ref{phi_eq_1}) splits into the two equations
\begin{align}
  \label{matrixSIT}
  \begin{split}
    p_{xy}=(q^2)_y \, ,\qquad
    q_{xy}=q-p_yq-q p_y \; .
  \end{split}
\end{align}
We refer to them as \emph{matrix-SIT} equations (see section~5), not 
purporting that they have a similar physical relevance as in the scalar case. 
The Miura transformation (\ref{Miura}) (with $\Delta=0$) now reads
\begin{align}    \label{Miura_matrixSIT}
  \begin{split}
    g_x \, g^{-1} = \tfrac{1}{2} \, [J,\phi] \, ,\qquad
    \tfrac{1}{2} \, [J,g] \, g^{-1} = \phi_y \; .
  \end{split}
\end{align}
Writing 
\begin{align*}
  g =
  \begin{pmatrix}
    a & b \\
    c & d \\
  \end{pmatrix} \, ,
\end{align*}
with $m \times m$ matrices $a,b,c,d$, and assuming that $a$ and its Schur complement 
$\mathcal{S}(a) = d - c \, a^{-1} b$ is invertible 
(which implies that $g$ is invertible), (\ref{Miura_matrixSIT}) with (\ref{phi_matrixSIT}) 
requires
\be  
     b = -c \, a^{-1} d \, , \qquad 
     a_x = - c_x \, a^{-1} c \, , \qquad
     d_x = -c_x \, a^{-1} c \, a^{-1} d \; .   \label{dual_matrixSIT_cond}
\ee
The last equation can be replaced by $d_x \, d^{-1} = a_x \, a^{-1}$. 
Invertibility of $\mathcal{S}(a)$ implies that $d$ and $I + r^2$ are invertible, where 
$r := c \, a^{-1}$. 
The conditions (\ref{dual_matrixSIT_cond}) are necessary in order 
that the Miura transformation relates solutions of (\ref{matrixSIT}) to solutions of its `dual'
\begin{align}  \label{g_eq_1}
           (g_x \, g^{-1})_y = \tfrac{1}{4} \, [gJg^{-1} , J] \, , 
\end{align}
obtained from (\ref{g_eq}). Taking (\ref{dual_matrixSIT_cond}) into account, the Miura transformation reads
\be
       q = -c_x \, a^{-1} = -r_x - r \, a_x \, a^{-1} \, , \quad
     q_y = - r \, (I + r^2)^{-1} \, , \quad
     p_y = I - (I + r^2)^{-1} \; .    \label{matrixSIT_Miura}
\ee
As a consequence, we have
 \begin{align}   \label{SITsol_id}
    {q_y}^2 + {p_y}^2 = p_y \; .
 \end{align}
Furthermore, the second of (\ref{dual_matrixSIT_cond}) and the first of (\ref{matrixSIT_Miura}) 
imply $a_x a^{-1} = q r$. Hence we obtain the system 
\be
     r_x = -q - r \, q \, r \, , \qquad
     q_y = - r \, ( I + r^2 )^{-1} \, ,   \label{ncsGsys}
\ee
which may be regarded as a matrix or `noncommutative' generalization of the 
sine-Gordon equation. There are various such 
generalizations in the literature. The first equation has the solution 
$q = - \sum_{k=0}^\infty (-1)^k \, r^k \, r_x \, r^k$, if the sum exists. 
Alternatively, we can express this as $q = -(I + r_L r_R)^{-1}(r_x)$, where
$r_L$ ($r_R$) denotes the map of left (right) multiplication by $r$. 
This can be used to eliminate $q$ from the second equation, resulting in 
\be
      \left( (I + r_L r_R)^{-1}(r_x) \right)_y = r \, (I+r^2)^{-1} \; .  \label{ncsG}
\ee
If $r = \tan(\theta/2) \, \boldsymbol{\pi}$ with a constant projection $\boldsymbol{\pi}$ (i.e. 
$\boldsymbol{\pi}^2 = \boldsymbol{\pi}$) and a function $\theta$, then (\ref{ncsG}) reduces 
to the sine-Gordon equation 
\be
     \theta_{xy} = \sin \theta \; .   \label{sG}
\ee

(\ref{ncsGsys}) can be obtained directly from (\ref{g_eq_1}) as follows, by setting
\begin{align*}
  g =
  \begin{pmatrix}
    a & -c \\
    c & a \\
  \end{pmatrix} 
= \begin{pmatrix}
    I & -r \\
    r & I \\
  \end{pmatrix} \, a \, , \quad \mbox{hence} \quad
   g^{-1} = a^{-1} \, 
  \begin{pmatrix}
    I & r \\
   -r & I \\
  \end{pmatrix} \, (I + r^2)^{-1} \; .
\end{align*}
This leads to
\bez
    \left( (r_x \, r + r \rho \, r + \rho)(I+r^2)^{-1} \right)_y = 0 \, , \quad
    \left( (r_x + r \rho - \rho \, r)(I+r^2)^{-1} \right)_y = r (I+r^2)^{-1} \, , 
\eez
where $\rho := a_x a^{-1}$. 
Setting an integration `constant' to zero, the first equation integrates to 
$\rho = - r_x r - r \rho \, r$. With its help, the second can be written as 
$(r_x + r \rho)_y = r (I+r^2)^{-1}$. Since $q = -(r a)_x \, a^{-1} = -r_x - r \, \rho$, this 
is the second of (\ref{ncsGsys}). The first follows noting that $q r = \rho$. 

\paragraph{5. Sharp line SIT equations and sine-Gordon.}
We consider the scalar case, i.e. $m=1$. Introducing $\mathcal{E} = 2 \sqrt{\alpha} q$ with a positive 
constant $\alpha$, $\mathcal{P} = 2 q_y$, $\mathcal{N} = 2 p_y -1$, and new coordinates 
$z,t$ via $x = \sqrt{\alpha} (z-t)$ and $y = \sqrt{\alpha} z$, the system (\ref{matrixSIT}) is transformed 
into
\bez
    \mathcal{P}_t = \mathcal{E} \, \mathcal{N} \, , \quad
    \mathcal{N}_t = -\mathcal{E} \, \mathcal{P} \, ,
\eez
and the relation between $\mathcal{E}$ and $\mathcal{P}$ takes the form 
\bez
    \mathcal{E}_z + \mathcal{E}_t = \alpha \, \mathcal{P} \; .
\eez
These are the sharp line self-induced transparency (SIT) equations \cite{Lamb71,CGEB73,BCEG74}. 
We note that $\mathcal{P}^2 + \mathcal{N}^2$ is conserved. Indeed, as a consequence of (\ref{SITsol_id}), 
we have $\mathcal{P}^2 + \mathcal{N}^2 =1$.  
Writing $\mathcal{P} = - \sin \theta$ and $\mathcal{N} = -\cos \theta$, reduces the first two equations 
to $\mathcal{E} = \theta_t$. Expressed in the coordinates $x,y$, the third then becomes the sine-Gordon equation (\ref{sG}) (cf. \cite{CGEB73}). As a consequence of the above relations, $q$ and $p$ depend as follows on $\theta$, 
\begin{gather}  \label{Miura_sG}
    q = -\tfrac{1}{2} \theta_x \, , \qquad
  q_y = - \tfrac{1}{2} \sin \theta \, , \qquad
  p_y = \tfrac{1}{2} (1-\cos \theta )    \; .
\end{gather}
These are precisely the equations that result from the Miura transformation (\ref{Miura_matrixSIT}) 
(or from (\ref{matrixSIT_Miura})), choosing 
\begin{align*}
  g=
  \begin{pmatrix}
    \cos{\tfrac{\theta}{2}}&-\sin{\tfrac{\theta}{2}} \\ \noalign{\medskip}
    \sin{\tfrac{\theta}{2}}&\cos{\tfrac{\theta}{2}}
  \end{pmatrix} \, ,
\end{align*}
and (\ref{g_eq_1}) becomes the sine-Gordon equation (\ref{sG}). 
The conditions (\ref{dual_matrixSIT_cond}) are identically satisfied as a consequence of the form of $g$.

\paragraph{6. A universal method of generating solutions from a matrix linear system.}
\begin{theorem}
\label{theorem:main}
Let $(\Omega, \d, \bar{\d})$ be a bidifferential calculus with
$\Omega = \cA \otimes \bigwedge(\bbC^2)$, where $\cA$ is the algebra of matrices with 
entries in some algebra $\cB$ (where the product of two matrices is defined to be zero if 
the sizes of the two matrices do not match). 
For fixed $N,N'$, let $\bX \in \mathrm{Mat}(N,N,\cB)$ and $\bY \in \mathrm{Mat}(N',N,\cB)$ 
be solutions of the linear equations
\bez
   \bar{\d} \bX = (\d \bX) \, \bP   \, , \qquad
   \bar{\d} \bY = (\d \bY) \, \bP \, , \qquad 
   \bR \, \bX - \bX \, \bP = -\bQ \, \bY     \, ,
\eez
with $\d$- and $\bd$-constant matrices $\bP,\bR \in \mathrm{Mat}(N,N,\cB)$, and 
$\bQ = \tbV \, \tbU$,  
where $\tbU \in \mathrm{Mat}(n,N',\cB)$ and $\tbV \in \mathrm{Mat}(N,n,\cB)$ are 
$\d$- and $\bar{\d}$-constant. 
If $\bX$ is invertible, the $n \times n$ matrix variable
\bez
     \phi = \tbU \bY \bX^{-1} \tbV  \in \mathrm{Mat}(n,n,\cB)    
\eez
solves $\bar{\d} \phi = (\d \phi) \, \phi + \d \vartheta$ with 
$\vartheta = \tbU \bY \bX^{-1} \bR \tbV$, hence (by application of $\d$) also (\ref{phi_eq}). \hfill  $\square$
\end{theorem}

There is a similar result for (\ref{g_mod_eq}) \cite{DMH10AKNS}. The Miura transformation 
is a corresponding bridge.

\paragraph{7. Solutions of the matrix SIT equations.}
 From Theorem~\ref{theorem:main} we can deduce the following result, using straightforward 
calculations \cite{Kanning10}, analogous to those in \cite{DMH10NLS} (see also \cite{DMH10AKNS}). 

\begin{proposition}
\label{prop:matrixSIT}
Let $\bS \in \mathrm{Mat}(M,M,\mathbb{C})$ be invertible, $\bU \in \mathrm{Mat}(m,M,\mathbb{C})$, 
  $\bV \in \mathrm{Mat}(M,m,\mathbb{C})$, and $\bK \in \mathrm{Mat}(M,M,\mathbb{C})$ a solution of 
the Sylvester equation
  \begin{align}
     \label{Sylvester}
    \bS \bK + \bK \bS = \bV \bU \; .
  \end{align}
Then, with $\bXi = e^{-\bS x-\bS^{-1} y}$ and any $p_0 \in \mathrm{Mat}(m,m,\mathbb{C})$ 
(more generally $x$-dependent), 
  \begin{align}
    \label{matrixSITsol}
    \begin{split}
      q = \bU \bXi \, (\bI_M + (\bK \bXi)^2)^{-1} \bV \, , \qquad
      p = p_0 - \bU \bXi \bK \bXi \, (\bI_M + (\bK \bXi)^2)^{-1} \bV 
    \end{split}
  \end{align}
(assuming the inverse exists) is a solution of (\ref{matrixSIT}). \hfill $\square$
\end{proposition}

If the matrix $\bS$ satisfies the spectrum condition
\begin{align}
   \label{spec_cond}
  \sigma(\bS) \cap \sigma(-\bS) = \emptyset 
\end{align}
(where $\sigma(\bS)$ denotes the set of eigenvalues of $\bS$), 
then the Sylvester equation (\ref{Sylvester}) has a unique solution $\bK$ (for any choice of the 
matrices $\bU, \bV$), see e.g. \cite{Horn+John91}. 

By a lengthy calculation \cite{Kanning10} one can verify directly that the solutions in 
Proposition~\ref{prop:matrixSIT} satisfy (\ref{SITsol_id}). Alternatively, one can show that 
these solutions actually determine solutions of the Miura transformation (cf. \cite{DMH10AKNS}), 
and we have seen that (\ref{SITsol_id}) is a consequence. 

There is a certain redundancy in the matrix data that determine the solutions (\ref{matrixSITsol})
of (\ref{matrixSIT}). This can be narrowed down by observing that the following 
transformations leave (\ref{Sylvester}) and (\ref{matrixSITsol}) invariant (see also the NLS case 
treated in \cite{DMH10NLS}). \\
(1) \emph{Similarity transformation} with an invertible $\bM \in \mathrm{Mat}(M,M,\mathbb{C})$ :
\begin{align*}
  \bS \mapsto \bM \bS \bM^{-1} \, , \quad
  \bK \mapsto \bM \bK \bM^{-1} \, , \quad
  \bV \mapsto \bM \bV \, , \quad
  \bU \mapsto \bU \bM^{-1} \; .
\end{align*}
As a consequence, we can choose $\bS$ in Jordan normal form without restriction of generality. \\
(2) \emph{Reparametrization transformation} with invertible $\bA, \bB \in \mathrm{Mat}(M,M,\mathbb{C})$ :
\begin{align*}
  \bS \mapsto \bS \, ,\quad
  \bK \mapsto \bB^{-1} \bK \bA^{-1} \, , \quad
  \bV \mapsto \bB^{-1} \bV \, , \quad
  \bU \mapsto \bU \bA^{-1} \, , \quad
  \bXi \mapsto \bA \bB \bXi \; .
\end{align*}
(3) \emph{Reflexion symmetry}:
\begin{gather*}
  \bS \mapsto -\bS \, , \quad
  \bK \mapsto -\bK^{-1} \, , \quad
  \bV \mapsto \bK^{-1} \bV \, , \quad
  \bU \mapsto \bU \bK^{-1} \, , \quad
  p_0 \mapsto p_0 -\bU \bK^{-1} \bV \; .
\end{gather*}
This requires that $\bK$ is invertible. More generally, such a reflexion can be applied to any 
Jordan block of $\bS$ and then changes the sign of its eigenvalue \cite{Kanning10} (see also \cite{ABDM10,DMH10NLS}). 
The Jordan normal form can be restored afterwards via a similarity transformation. 

The following result is easily verified \cite{Kanning10}.

\begin{proposition}
\label{prop:herm_red}
Let $\bS, \bU, \bV$ be as in Proposition~\ref{prop:matrixSIT} and $\bT \in \mathrm{Mat}(M,M,\mathbb{C})$ invertible. \\
(1) Let $\bT$ be Hermitian (i.e. $\bT^\dagger = \bT$) and such that $\bS^\dagger = \bT \bS \bT^{-1}$, 
$\bU = \bV^\dagger \bT$. Let $\bK$ be a solution of (\ref{Sylvester}), which can then be chosen such that
$\bK^\dagger = \bT \bK \bT^{-1}$. Then $q$ and $p$ given by (\ref{matrixSITsol}) with 
$p_0^\dagger = p_0$ are both Hermitian and thus solve the Hermitian reduction of (\ref{matrixSIT}). \\
(2) Let $\bar{\bT} = \bT^{-1}$ (where the bar means complex conjugation) and 
$\bar{\bS} = \bT \bS \bT^{-1}$, $\bar{\bU} = \bU \bT^{-1}$ and $\bar{\bV} = \bT \bV$. Let $\bK$ be a solution of (\ref{Sylvester}), which can then be chosen such that
$\bar{\bK} = \bT \bK \bT^{-1}$. Then $q$ and $p$ given by (\ref{matrixSITsol}) with 
$\bar{p}_0 = p_0$ satisfy $\bar{q} = q$ and $\bar{p}=p$, and thus solve the complex 
conjugation reduction of (\ref{matrixSIT}). 
\hfill $\square$
\end{proposition}

\paragraph{8. Rank one solutions.}
Let $M=1$. We write $\bS =s$, $\bU = \bu$, 
$\bV = \bv^\intercal$, $\bK = k$ (where ${}^\intercal$ means the transpose)
and $\bXi = \xi = e^{-s x - s^{-1} y}$. Then (\ref{Sylvester}) 
yields $k = ( \bv^\intercal \bu )/(2 s)$. From (\ref{matrixSITsol}) we obtain
\bez
    q = \frac{2 \, s \, k \, \xi}{1 + (k \xi)^2} \, \boldsymbol{\pi} \, , \quad
    p = \tilde{p}_0 + \frac{2 \, s }{1 + (k \xi)^2} \, \boldsymbol{\pi} 
          \, , \quad
    \tilde{p}_0 := p_0 - 2 s \, \boldsymbol{\pi} \, , \quad
    \boldsymbol{\pi} := \frac{\bu \bv^\intercal}{\bv^\intercal \bu} \; .
\eez
The Miura transformation (\ref{matrixSIT_Miura}) implies $r = - q_y \, (I-p_y)^{-1}$, and we obtain
\bez
    r = - \frac{2 \, k \xi}{1 - (k \xi)^2} \, \boldsymbol{\pi} \, ,
\eez
which is singular. But $\theta = -2 \arctan(2 k \xi/[1 - (k \xi)^2])$ is the single kink 
solution of the sine-Gordon equation (\ref{sG}).

\begin{figure}[H] 
\begin{center} 
\begin{minipage}{5cm}
\resizebox{!}{3.cm}{
\includegraphics{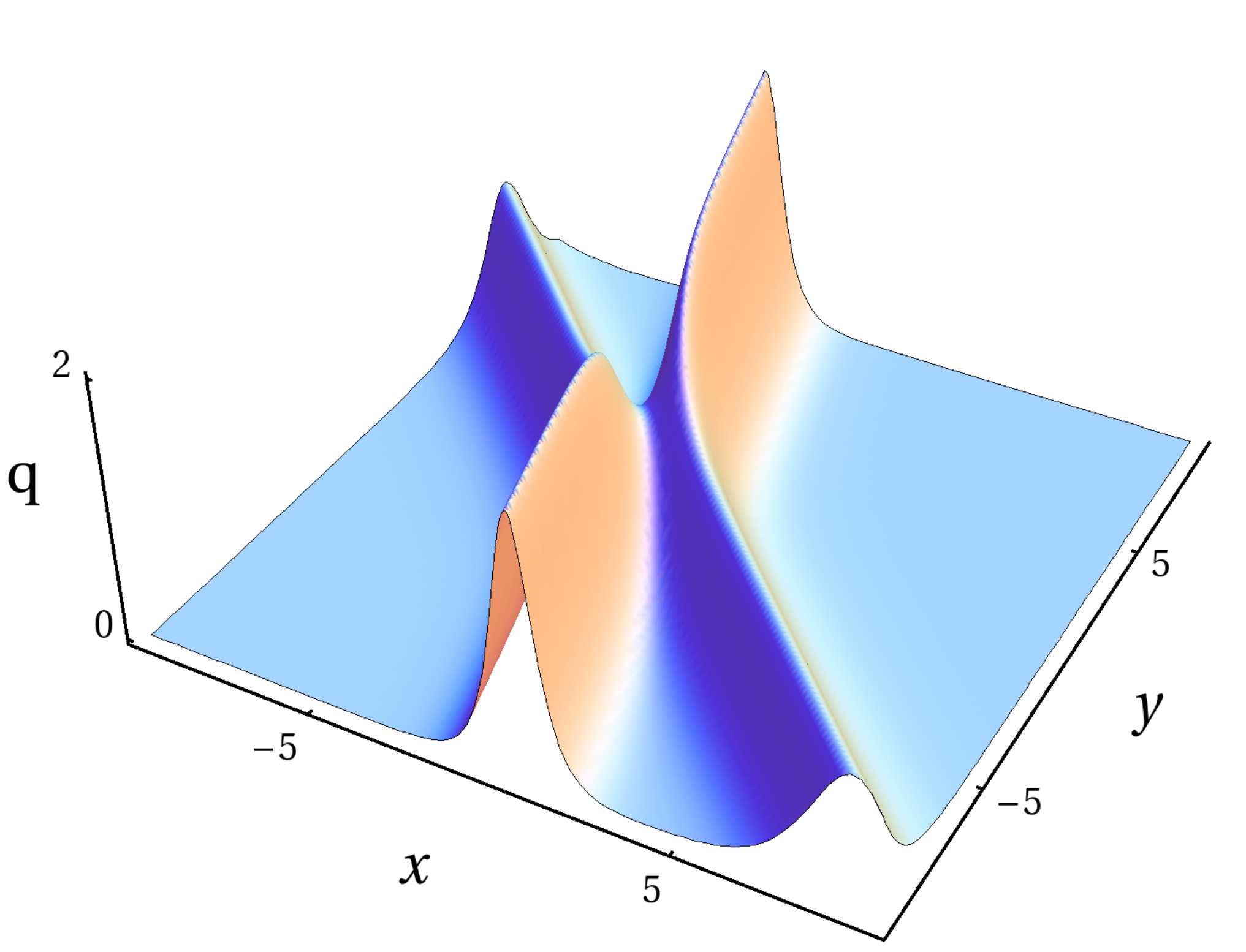}
}
\end{minipage}
\hspace{.5cm}
\begin{minipage}{5cm}
\vspace{-.3cm}
\resizebox{!}{3.cm}{
\includegraphics{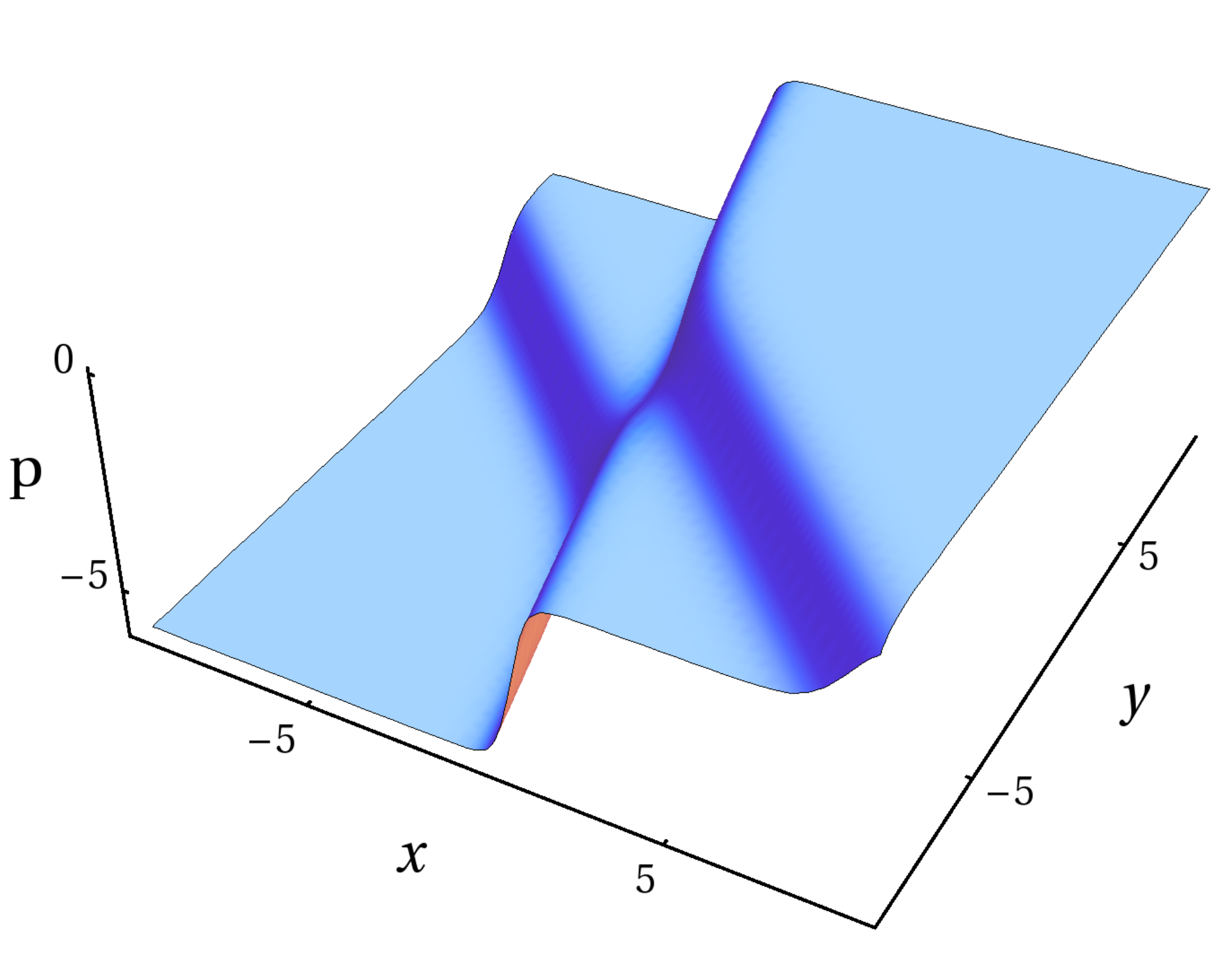}
}
\end{minipage}
\parbox{15cm}{
\caption{A scalar 2-soliton solution with $\bS = \mathrm{diag}(1,2)$ and $u_i =v_i =1$. 
\label{fig:2soliton}  }
}
\end{center}
\end{figure}

\begin{figure}[H] 
\begin{center} 
\begin{minipage}{5cm}
\resizebox{!}{3.cm}{
\includegraphics{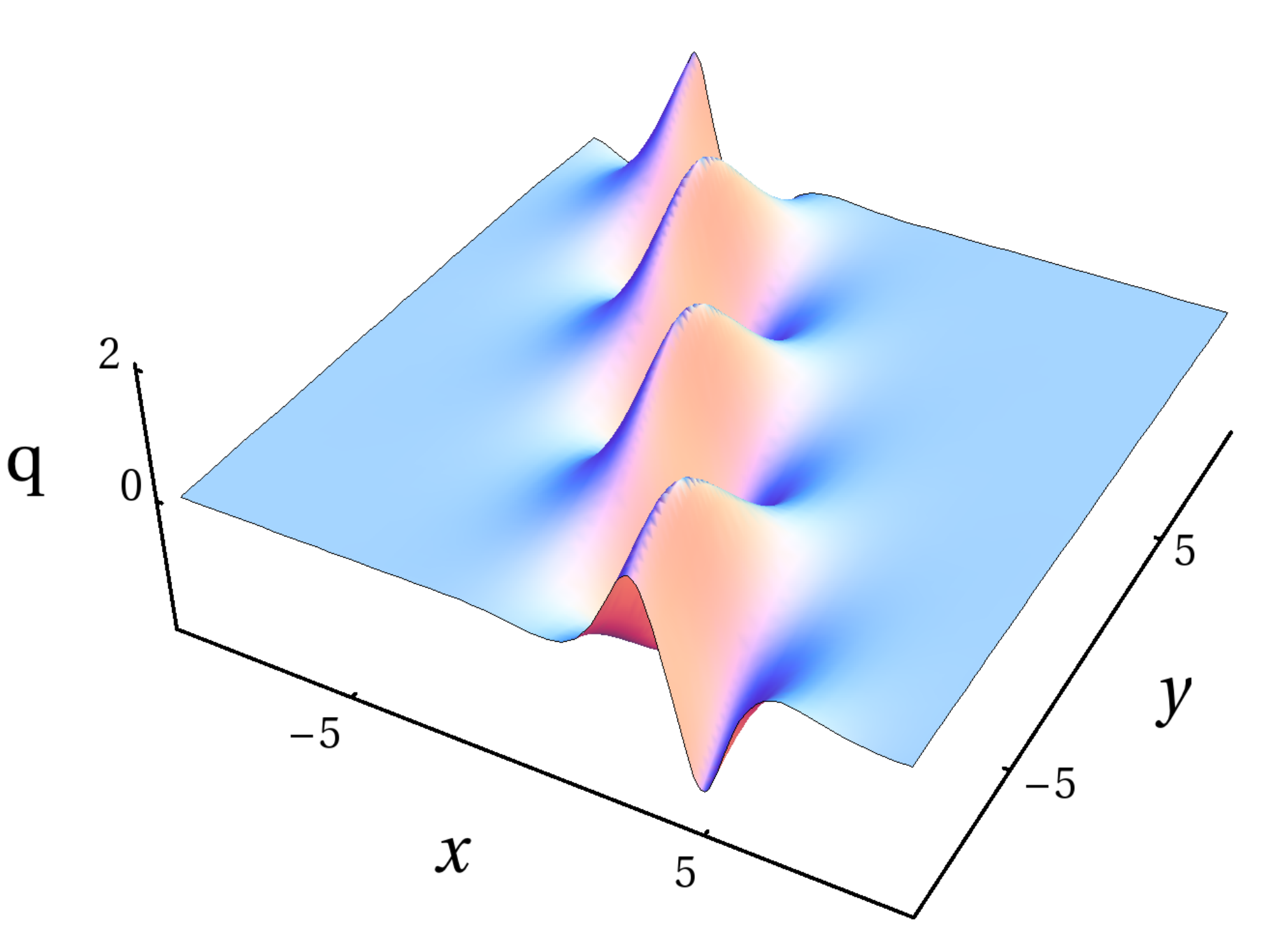}
}
\end{minipage}
\hspace{.5cm}
\begin{minipage}{5cm}
\vspace{-.3cm}
\resizebox{!}{3.cm}{
\includegraphics{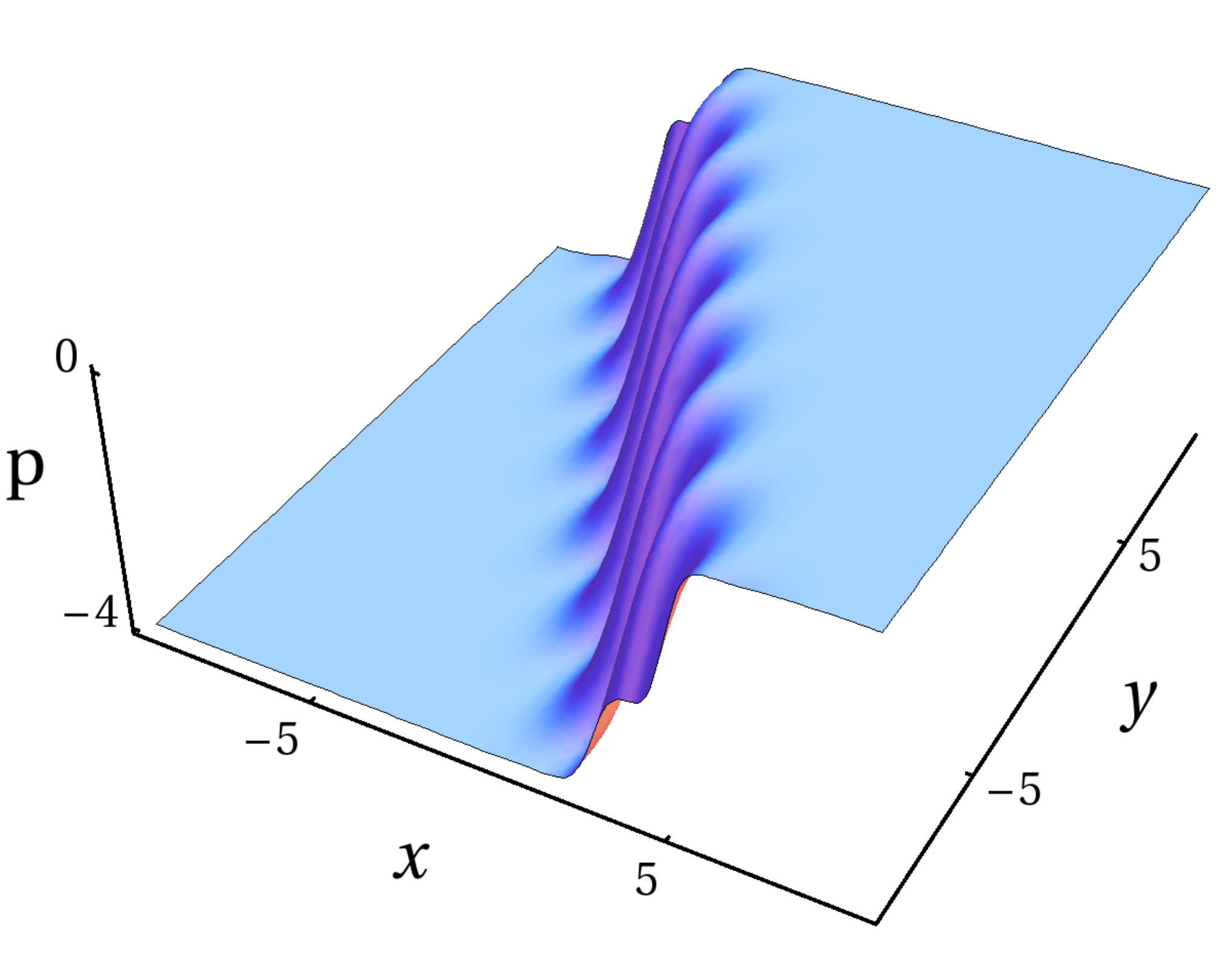}
}
\end{minipage}
\parbox{15cm}{
\caption{A scalar breather solution with $\bS = \mathrm{diag}(1+\mathrm{i},1-\mathrm{i})$ 
and $u_i =v_i =1$. \label{fig:breather}  }
}
\end{center}
\end{figure}

\paragraph{9. Solutions of the scalar (sharp line) SIT equations.}
We rewrite $p$ in (\ref{matrixSITsol}), where now $m=1$, as follows,
 \begin{align}
   p&= p_0 - \mathrm{tr} \left((\bS \bK + \bK \bS) \, \bXi \bK \bXi \, (\bI_M + (\bK \bXi)^2)^{-1}\right) 
       \nonumber \\
    &= p_0 + \mathrm{tr} \left((\bI_M + (\bK \bXi)^2)_x \, (\bI_M + (\bK \bXi)^2)^{-1}\right) \nonumber \\
    &= p_0 + \left(\log \det \left(\bI_M + (\bK \bXi)^2\right) \right)_x   \, ,  \label{SIT_p}
  \end{align}
using (\ref{Sylvester}) and the identity $(\det\bM)_x = \mathrm{tr} (\bM_x \bM^{-1}) \, \det \bM$ 
for an invertible matrix function $\bM$. $q$ in (\ref{matrixSITsol}) can be expressed as
\begin{align*}
   q = 2 \, \mathrm{tr}\left( \bS \bK \bXi \, (\bI_M + (\bK \bXi)^2)^{-1}  \right) \; .
 \end{align*}
In particular, if $\bS$ is diagonal with eigenvalues $s_i$, $i=1,\ldots,M$, and satisfies (\ref{spec_cond}), 
then the solution $\bK$ of the Sylvester equation (\ref{Sylvester}), which now amounts to 
$\mathrm{rank}(\bS \bK + \bK \bS) =1$, is the Cauchy-type matrix with components
$K_{ij} = v_i \, u_j /(s_i+s_j)$, where $u_i,v_i \in \mathbb{C}$. Figs.~\ref{fig:2soliton} 
and \ref{fig:breather} show plots of two examples from the above family of solutions. 

\paragraph{10. A family of solutions of the real sine-Gordon equation.}
Via the Miura transformation (\ref{Miura_sG}), 
Proposition~\ref{prop:matrixSIT} determines a family of sine-Gordon solutions (see also 
e.g. \cite{CGEB73,Hiro72,AKNS73sG,Poeppe83,Zheng86,Schie09,ADM10} 
for related results obtained by different methods).

\begin{proposition}
\label{prop:sG_sol}
Let $\bS \in \mathrm{Mat}(M,M,\mathbb{C})$ be invertible and $\bK \in \mathrm{Mat}(M,M,\mathbb{C})$ 
such that $\mathrm{rank}(\bS \bK + \bK \bS) =1$, $\det(\bI_M + (\bK \bXi)^2) \in \mathbb{R}$ 
with $\bXi = e^{-\bS x-\bS^{-1} y}$, 
and $\mathrm{tr}( \bS \bK \bXi \, (\bI_M + (\bK \bXi)^2)^{-1}) \not\in \mathrm{i} \mathbb{R}$ 
(where $\mathrm{i}$ is the imaginary unit). 
Then 
  \begin{align}   \label{sG_sol}
    \theta = 4 \arctan \Big( \frac{\sqrt{\beta}}{1+\sqrt{1-\beta}} \Big) \quad
      \mbox{with} \quad \beta := \left(\log |\det(\bI_M + (\bK \bXi)^2)| \right)_{xy} 
  \end{align}
solves the sine-Gordon equation $\theta_{xy} = \sin \theta$ in any open set of $\mathbb{R}^2$ where
$\det(\bI_M + (\bK \bXi)^2) \neq 0$. 
\end{proposition}
\noindent
\textit{Proof:} Let $p$ be given by (\ref{SIT_p}). Due to the assumption $\det(\bI_M + (\bK \bXi)^2) \in \mathbb{R}$, 
$p_y$ is real, hence (\ref{SITsol_id}) implies $|1-2 p_y|^2 = 1 - 4 {q_y}^2$. It follows that ${q_y}^2$ is real. 
Since another of our assumptions excludes that $q_y$ is imaginary, it follows that $|1-2 p_y| \leq 1$.
Hence the equation $\cos \theta = 1-2 p_y$ (second of (\ref{Miura_sG})) has a real solution $\theta$. 
Inserting the expression (\ref{SIT_p}) for $p$, we arrive at 
$\cos \theta = 1-2 \left(\log \det(\bI_M + (\bK \bXi)^2) \right)_{xy}$.
Moreover, (\ref{SITsol_id}) shows that $p_y \geq 0$ and thus $0 \leq p_y \leq 1$. 
Using identities for the inverse trigonometric functions, we find (\ref{sG_sol}), where $\beta = p_y$. 
\hfill $\square$
\vskip.2cm

Proposition~\ref{prop:herm_red} yields sufficient conditions on the matrix data for which 
the last two assumptions in Proposition~\ref{prop:sG_sol} are satisfied.


\small

\begin{thebibliography}{10}

\bibitem{DMH08bidiff}
Dimakis, A., M\"uller-Hoissen, F.: \textit{Bidifferential graded algebras and
  integrable systems}. Discr. Cont. Dyn. Systems Suppl., {\bf 2009}, 2009, p. 208--219.

\bibitem{DMH10NLS}
Dimakis, A., M\"uller-Hoissen, F.: \textit{Solutions of matrix {NLS} systems and
  their discretizations: a unified treatment}. Inverse Problems, {\bf 26}, 2010, 
  095007.

\bibitem{DMH10AKNS}
Dimakis, A., M\"uller-Hoissen, F.:  \textit{Bidifferential calculus approach to
  {AKNS} hierarchies and their solutions}. SIGMA, {\bf 6}, 2010, 2010055. 

\bibitem{Gris+Pena03}
Grisaru, M., Penati, S.: \textit{An integrable noncommutative version of the
  sine{-G}ordon system}. Nucl. Phys. B, {\bf 655}, 2003, p. 250--276.

\bibitem{Lamb71}
Lamb, G.: \textit{Analytical descriptions of ultrashort optical pulse propagation in
  a resonant medium}. Rev. Mod. Phys., {\bf 43}, 1971, p. 99--124.

\bibitem{CGEB73}
Caudrey, P., Gibbon, J., Eilbeck, J., Bullough, R.: \textit{Exact multi-soliton
  solutions of the self-induced transparency and sine-{G}ordon equations}.
  Phys. Rev. Lett., {\bf 30}, 1973, p. 237--238.

\bibitem{BCEG74}
Bullough, R., Caudrey, P., Eilbeck, J., Gibbon, J.:  \textit{A general theory of
  self-induced transparency}. Opto-electronics, {\bf 6}, 1974, p. 121--140.

\bibitem{Kanning10}
Kanning, N.: \textit{Integrable Systeme in der Allgemeinen
  Relativit\"atstheorie: ein Bidifferentialkalk\"ul-Zugang}. Diploma thesis. 
  G\"ottingen: University of G\"ottingen, 2010. 

\bibitem{Horn+John91}
Horn, R., Johnson, C.: \textit{Topics in Matrix Analysis}. Cambridge:
  Cambridge Univ. Press, 1991.

\bibitem{ABDM10}
Aktosun, T., Busse, T., Demontis, F., van~der Mee, C.: \textit{Symmetries for exact
  solutions to the nonlinear {S}chr\"odinger equation}. J. Phys. A: Theor.
  Math., {\bf 43}, 2010, 025202.

\bibitem{Hiro72}
Hirota, R.: \textit{Exact solution of the sine-{G}ordon equation for multiple
  collisions of solitons}. J. Phys. Soc. Japan, {\bf 33}, 1972, p. 1459--1463.

\bibitem{AKNS73sG}
Ablowitz, M., Kaup, D., Newell, A., Segur, H.: \textit{Method for solving the
  sine-{G}ordon equation}. Phys. Rev. Lett., {\bf 30}, 1973, p. 1262--1264.

\bibitem{Poeppe83}
P\"oppe, C.: \textit{Construction of solutions of the sine-{G}ordon equation by means
  of {F}redholm determinants}. Physica D, {\bf 9}, 1983, p. 103--139.

\bibitem{Zheng86}
Zheng, W.: \textit{The sine-{G}ordon equation and the trace method}. J. Phys. A:
  Math. Gen., {\bf 19}, 1986, p. L485--L489.

\bibitem{Schie09}
Schiebold, C.: \textit{Noncommutative {AKNS} system and multisoliton solutions to the
  matrix sine-{G}ordon equation}. Discr. Cont. Dyn. Systems Suppl., {\bf 2009}, 
  2009, p. 678--690.

\bibitem{ADM10}
Aktosun, T., Demontis, F., van~der Mee, C.: \textit{Exact solutions to the
  sine-{G}ordon equation}. arXiv:1003.2453, 2010.

\end{thebibliography}
\end{document}